\documentclass[twocolumn,aps,pra,showpacs,longbibliography,superscriptaddress]{revtex4-1}
\usepackage{amssymb,latexsym}
\usepackage{mathrsfs}
\usepackage{lipsum}
\usepackage{graphicx}
\usepackage[caption=false]{subfig}
\usepackage{mathtools}
\usepackage{epstopdf}
\usepackage{xcolor}
\usepackage[colorlinks, linkcolor=red, anchorcolor=green, citecolor=green, urlcolor=blue]{hyperref}
\DeclareGraphicsExtensions{.pdf,.jpg,.png,.eps}
\usepackage[toc,page,title,titletoc,header]{appendix}
\usepackage{etoolbox}
\usepackage{breqn}

\makeatletter
\let\cat@comma@active\@empty
\makeatother

\begin{document}
	
\title{Phase sensitivity approaching quantum Cram\'er-Rao bound in a modified SU(1,1) interferometer}
	
\author{Jian-Dong Zhang}
\email[]{zhangjiandong1993@gmail.com}
\affiliation{School of Mathematics and Physics, Jiangsu University of Technology, Changzhou 213001, China}
\author{Chenglong You}
\affiliation{Quantum Photonics Laboratory, Department of Physics \& Astronomy, Louisiana State University, Baton Rouge, LA 70803, USA}
\author{Chuang Li}
\affiliation{Research Center for Quantum Sensing, Intelligent Perception Research Institute, Zhejiang Lab, Hangzhou 310000, China}
\author{Shuai Wang}
\affiliation{School of Mathematics and Physics, Jiangsu University of Technology, Changzhou 213001, China}

\date{\today}
	
\begin{abstract} 
SU(1,1) interferometers, based on the usage of nonlinear elements, are superior to passive interferometers in phase sensitivity. However, the SU(1,1) interferometer cannot make full use of photons carrying phase information as the second nonlinear element annihilates some of the photons inside. Here, we focus on improving phase sensitivity and propose a new protocol based on a modified SU(1,1) interferometer, where the second nonlinear element is replaced by a beam splitter. We utilize two coherent states as inputs and implement balanced homodyne measurement at the output. Our analysis suggests that the protocol we propose can achieve sub-shot-noise-limited phase sensitivity and is robust against photon loss and background noise. Our work is important for practical quantum metrology using SU(1,1) interferometers. 
\end{abstract}
	
\maketitle
	
\section{Introduction}
Phase estimation is an important means of precision measurement, which provides a route to estimate many physical quantities that cannot be directly measured via conventional methods. 
Related to this, optical interferometers are an effective tool for phase estimation.
In general, a conventional interferometer utilizes a coherent light source and intensity-based measurement; as a result, the phase sensitivity is bounded by the shot-noise limit $\Delta {\theta _{\rm{SNL}}} = {1/ {\sqrt {N} } }$ with $N$ photons inside the interferometer on average \cite{PhysRevD.23.1693}. To improve the phase sensitivity, two kinds of approaches are proposed. The first approach is to deploy quantum resources, including exotic non-classical states \cite{PhysRevLett.100.073601,PhysRevLett.104.103602,PhysRevLett.107.083601,PhysRevLett.110.163604,PhysRevLett.112.103604} and photon-number-resolving measurements \cite{Plick_2010,PhysRevA.87.043833,you2020multiphoton}; the second one is to change the evolution process through the use of nonlinear elements. A typical interferometer designed in terms of the second approach is usually known as an SU(1,1) interferometer, introduced by Yurke \emph{et al}. \cite{PhysRevA.33.4033}, where two beam splitters in the conventional interferometer are replaced by two optical parametric amplifiers (OPAs).	

In recent years, a large number of theoretical and experimental studies have focused on phase estimation using SU(1,1) interferometers.
Plick \emph{et al}. demonstrated that an SU(1,1) interferometer with two coherent states as inputs can achieve the sub-shot-noise-limited phase sensitivity by using conventional intensity measurement \cite{Plick2010}.
Li \emph{et al}. proposed an SU(1,1) interferometer fed by coherent and squeezed vacuum states, the phase sensitivity can approach the Heisenberg limit with balanced homodyne measurement \cite{Li_2014} and parity measurement \cite{PhysRevA.94.063840}.
An SU(1,1) interferometer using squeezed coherent state and coherent state was presented by Hu \emph{et al}., they studied the phase sensitivity with balanced homodyne measurement and discussed the optimal squeezing fraction \cite{Hu2016}.
Meanwhile, there has been an interest in modified SU(1,1) interferometers, such as pumped-up and truncated SU(1,1) interferometers \cite{PhysRevLett.118.150401,Anderson:17,PhysRevA.95.063843}.
In addition, the phase sensitivity limit of an SU(1,1) interferometer was studied.
Gong \emph{et al}. analyzed the quantum Cram\'er-Rao bounds of various Gaussian inputs \cite{:94205}. More recently, You \emph{et al}. discussed the conclusive sensitivity limit for a protocol using single-mode inputs.  \cite{PhysRevA.99.042122}.

Indeed, the phase sensitivities of most protocols cannot achieve or approach the corresponding quantum Cram\'er-Rao bound, although they are capable of reaching the Heisenberg limit \cite{:94205}.
A major reason for this result originates from the drawback of SU(1,1) interferometers: many photons stimulated by the first OPA are absorbed by the second OPA. 
Therefore, only a part of photons carrying phase information contribute to the measurement, which limits the phase sensitivity.
In this paper, we present a new protocol for phase estimation using a modified SU(1,1) interferometer; the second OPA in the conventional SU(1,1) interferometer is replaced by a beam splitter.
Two coherent states are used as inputs, and balanced homodyne measurement is implemented at the output.
The phase sensitivity of our protocol is compared with the previous SU(1,1)-type protocol and its corresponding quantum Cram\'er-Rao bound.
For practical purposes, we also analyze the effects of photon loss and background noise on phase sensitivity of our protocol. 

The paper is organized as follows.
Sec. \ref{II} introduces model of the modified SU(1,1) interferometer and analyzes the phase sensitivity in an ideal scenario.
In Sec. \ref{III}, we compare the phase sensitivity of our protocol with that of the previous SU(1,1)-type protocol, and analyze the optimality of our protocol by calculating the quantum Cram\'er-Rao bound.
The effects of photon loss and background noise on phase sensitivity are discussed in Sec. \ref{IV}.	
Finally, we summarize our results in Sec. \ref{V}.
	
\section{Model and phase sensitivity of modified SU(1,1) interferometer}
\label{II}
\subsection{Model}
\begin{figure}[htbp]
	\centering
	\includegraphics[width=0.48\textwidth]{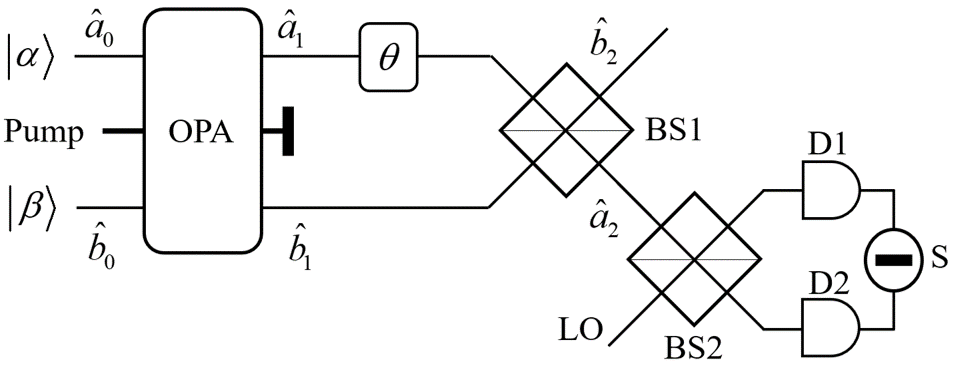}
	\caption{Schematic of phase estimation protocol based on a modified SU(1,1) interferometer.
	A beam splitter, BS1, takes the place of the second OPA in a conventional SU(1,1) interferometer. 
	One of the paths is then measured using balanced homodyne detection. 
	Here the transmittance of BS1 and BS2 is given by $\eta$ and 1/2, respectively. 
	The creation and annihilation operators for each stage are shown and given in the Appendix B. 
	OPA, optical parametric amplifier; BS, beam splitter; LO, local oscillator; D, detector; S, subtracter.}
	\label{system}
\end{figure}

The schematic of the modified SU(1,1) interferometer is shown in Fig. \ref{system}.
Two coherent states, $\left| \alpha  \right\rangle $ and $\left| \beta  \right\rangle$, are injected into a pumped OPA. Then the generated states undergo phase shift of $\theta$ on one of the paths. 
After the unknown phase shift, these states are recombined by a beam splitter with transmittance of $\eta$. Finally, balanced homodyne measurement is implemented at one of the outputs.

We assume that phase and gain parameter of the OPA are 0 and $g$, respectively; accordingly, the complex amplitudes of two coherent states can be expressed as $\alpha  = \left| \alpha  \right|{e^{i\varphi }}$ and $\beta  = \left| \beta  \right|{e^{i\delta }}$.
Since the OPA changes the mean photon number of the input state, we use the mean photon number inside our interferometer as the benchmark. 
Thus the mean photon number is given by
\begin{align}
\nonumber {N} = {\kern 0.5pt} & \langle {\hat a_1^\dag \hat a_1^{} + \hat b_1^\dag \hat b_1^{}} \rangle  \\
\nonumber = {\kern 1pt}{\kern 1pt} & ( {{{\left| \alpha  \right|}^2} + {{\left| \beta  \right|}^2}} ){\cosh }\left( {2g} \right) + 2{\sinh ^2}g \\
& + 2\left| {\alpha \beta } \right|\sinh \left( {2g} \right)\cos \left( {\varphi  + \delta } \right).
\label{e1}
\end{align} 
It is straightforward to show that when the total mean photon number of the input ${N}_{\text{in}}=\left| \alpha  \right|^2 + \left| \beta  \right|^2$ is constant, $ \left| \alpha  \right| = \left| \beta  \right| $ and $\varphi  + \delta = 0$ are the optimal conditions to obtain maximal mean photon number ${N}_{\text{max}}$ in our protocol. Throughout this paper, we assume such conditions are always met. 
That is, the mean photon number in our protocol is
\begin{align}
{N} = 2{\left| \alpha  \right|^2}{e^{2g}} + 2{\sinh ^2}g.
\label{e2}
\end{align}

\subsection{Phase sensitivity}
The measurement strategy in our protocol is balanced homodyne measurement, which can measure the quadrature of output by mixing with a strong coherent state.
The phase therefore can be estimated from the difference between the two detectors D1 and D2.
Here we consider amplitude quadrature $\hat X$ as measured observable, and the measurement operator can be written as
\begin{equation}
\hat X = \hat a_2^\dag  + \hat a_2^{}.
\label{e3}
\end{equation}

The creation operator of the output modes is given by (See Appendix \ref{B})
\begin{align}
\nonumber \hat a_2^\dag  = {\kern 1pt}{\kern 1pt} & \sqrt \eta  {e^{ - i\theta }}( {\hat a_0^\dag \cosh g + \hat b_0^{}\sinh g} ) \\
&- i\sqrt {1 - \eta } ( {\hat b_0^\dag \cosh g + \hat a_0^{}\sinh g} ).
\label{e4}
\end{align}

By using Eqs. (\ref{e3}) and (\ref{e4}), the expectation value of measurement operator is found to be
\begin{equation}
\langle {\hat X} \rangle  = 2{e^g}\sqrt \eta  \left| {\alpha } \right| \cos \theta.
\label{e5}
\end{equation}
Using the definition of visibility \cite{doi:10.1080/00107510802091298},
\begin{equation}
	{\cal V} = \frac{{{{\langle \hat X \rangle }_{\max }} - {{\langle \hat X \rangle }_{\min }}}}{{| {{{\langle \hat X \rangle }_{\max }}} | + | {{{\langle \hat X \rangle }_{\min }}} |}},
	\label{e5-2}
\end{equation}
we can find that the visibility of $\langle {\hat X} \rangle$ is 100\% for any transmittance.

Furthermore, we can calculate that
\begin{align}
\nonumber \langle {{{\hat X}^2}} \rangle  = {\kern 1pt}{\kern 1pt} & 2\eta {\left| \alpha  \right|^2}{e^{2g}}\left[ {\cos \left( {2\theta } \right) + 1} \right] + \cosh \left( {2g} \right) \\
&- 2\sinh \left( {2g} \right)\sqrt {\eta \left( {1 - \eta } \right)} \sin \theta.  
\label{e6}
\end{align}
	
The phase sensitivity of our protocol can be calculated via error propagation formula, defined as \cite{PhysRevA.93.023810}
\begin{equation}
\Delta \theta  = \frac{{\sqrt {\langle {{{\hat X}^2}} \rangle  - {{\langle {\hat X} \rangle ^2}}} }}{| {{{\partial {\langle {\hat X} \rangle}} \mathord{\left/{\vphantom {{\partial X} {\partial \theta }}} \right.\kern-\nulldelimiterspace} {\partial \theta }}} |}.
\label{e7}
\end{equation}
Plugging Eqs. (\ref{e5}) and (\ref{e6}) into Eq. (\ref{e7}), we get
\begin{equation}
\Delta \theta  = \frac{{\sqrt {\cosh \left( {2g} \right) - 2\sqrt {\eta \left( {1 - \eta } \right)} \sinh \left( {2g} \right)\sin \theta } }}{{2\sqrt \eta  {e^g}\left| {\alpha \sin \theta } \right|}}.
\label{e8}
\end{equation}
Clearly, the phase sensitivity depends on the phase shift $\theta$. Evidently, we obtain the optimal phase sensitivity at $\theta = \pi/2$.

Additionally, from Eq. (\ref{e8}), the phase sensitivity also depends on the value of transmittance $\eta$. We can obtain the optimal transmittance $\eta_{\rm{opt}}$ by taking the derivative of Eq. (\ref{e8}) with respect to $\eta$. When a fixed gain parameter $g$, the optimal transmittance $\eta_{\rm{opt}}$ is given by 
\begin{equation}
{\eta _{{\rm{opt}}}} = \frac{{{{\cosh }^2}\left( {2g} \right)}}{{{{\sinh }^2}\left( {2g} \right) + {{\cosh }^2}\left( {2g} \right)}}=\frac{1+\text{sech}(4g)}{2}.
\label{eq10}
\end{equation}

\begin{figure}[!htbp]
	\centering
	\includegraphics[width=0.48\textwidth]{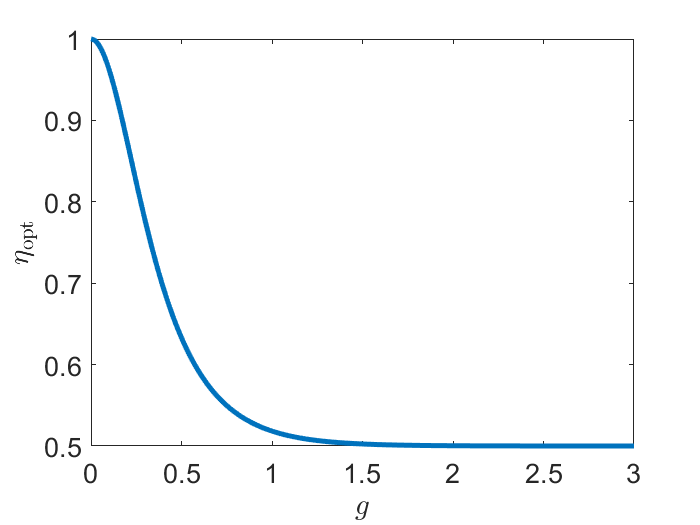}
	\caption{The optimal transmittance $\eta_{\rm{opt}}$ of the beam splitter as a function of gain parameter $g$.}
	\label{optimal}
\end{figure}

This reveals that the optimal transmittance depends solely on the gain parameter $g$, regardless of the mean photon number of the input.
If the gain parameter is known, then the transmittance $\eta$ can be chosen such that the phase sensitivity reaches its minimum. We note that this effect is pronounced at low gain parameters.
In Fig. \ref{optimal}, we illustrate the relationship between the optimal transmittance $\eta_{\rm{opt}}$ and gain parameter $g$. 
From Eq. (\ref{eq10}) and Fig. \ref{optimal}, we conclude that the optimal transmittance approaches 0.5 when $g$ is large. This suggests that a 50-50 beam splitter is nearly optimal for our protocol with a high gain parameter. Trivially, when $g = 0$, our protocol reduces to a classical interferometer, which is consistent with the result reported in Ref. \cite{zhang2020classical}. 

In the following sections, the transmittance is fixed to be $1/2$ for the sake of simplicity. 
Related to this, the optimal phase sensitivity turns out to be
\begin{equation}
\Delta {\theta _{\min }} = \frac{1}{{\sqrt 2 {e^{2g}}\left| \alpha  \right|}}.
\label{e9}
\end{equation}

According to the definition of mean photon number in Eq. (\ref{e2}), the shot-noise limit can be expressed as 
\begin{align}
\Delta {\theta _{\rm{SNL}}} = \frac{1}{{\sqrt {2{{\left| \alpha  \right|}^2}{e^{2g}} + 2{{\sinh }^2}g} }}.
\label{e10}
\end{align}
In general, high-intensity coherent states ($\left| \alpha  \right|^2  \gg 1 $) will be used as inputs, i.e., $ {{{\left| \alpha  \right|}^2}{e^{2g}} \gg {{\sinh }^2}g}$.
Then the shot-noise limit is approximately recast as
\begin{align}
\Delta {\theta _{\rm{SNL}}}  \simeq  \frac{1}{{\sqrt 2 {e^{g}{{{\left| \alpha  \right|}}} } }}.
\label{e11}
\end{align}

By comparing Eqs. (\ref{e9}) and (\ref{e11}), one can find that the phase sensitivity of our protocol is superior to the shot-noise limit by a factor of $e^g$.

\section{Comparison with the SU(1,1)-type protocol and the quantum Cram\'er-Rao bound}
\label{III}
In the previous section, we calculated the phase sensitivity of our protocol. In this section, we analyze the advantage of our protocol over the previous SU(1,1)-type protocol.
Then we compare the phase sensitivity with the quantum Cram\'er-Rao bound.

\subsection{Comparison with the SU(1,1)-type protocol}

In Ref. \cite{PhysRevA.94.063840}, Li \emph{et al}. used different measurement strategies to analyze the phase sensitivity of an SU(1,1)-type protocol with two equal coherent states as inputs. 
Their results indicate that balanced homodyne measurement gives the optimal phase sensitivity, which has the following form

\begin{align}
\Delta {\theta _{\rm{SU(1,1)}}} = \frac{1}{{2e^g\left| \alpha  \right|\sinh {g}}}.
\label{e12}
\end{align}

By comparing Eqs. (\ref{e9}) with (\ref{e12}), one can find that the phase sensitivity of our protocol outperforms the previous SU(1,1)-type protocol. To quantify the improvement, we calculate the improvement factor ${\cal K}$, given by
\begin{align}
{\cal K} = \frac{{\sqrt 2 {e^g}}}{{{e^g} - {e^{ - g}}}}.
\end{align}
 
For a high gain parameter, the phase sensitivity is improved by a factor of $\sqrt 2$.
With the decrease of gain parameter, the improvement factor of phase sensitivity increases rapidly. 
That is, our protocol can give at least a factor of $\sqrt 2$ improvement in phase sensitivity in comparison with an SU(1,1)-type protocol. 

In addition, noticing that the complexity of the measurement system is simplified.
The modified SU(1,1) interferometer circumvents problems encountered by the SU(1,1) interferometer, such as pump phase-matching and gain-matching \cite{Li_2014}.
Hence, our protocol remains superior in terms of sensitivity and complexity when compared with an SU(1,1)-type protocol.

\subsection{Comparison with the quantum Cram\'er-Rao bound}

Now we focus on the analysis of phase sensitivity limit of our protocol.
The ultimate limit of phase sensitivity is given by the quantum Cram\'er-Rao bound, $\Delta \theta_{\rm{QCRB}} = 1/{\sqrt {\cal F} }$ with $\cal F$ being the QFI.
In Ref. \cite{:94205}, Gong \emph{et al}. demonstrated that three different phase configurations would give three different QFIs with two coherent states as inputs.
This is due to the asymmetric gain of the OPA; in other words, we have $\langle {\hat a_1^\dag \hat a_1^{} \rangle \ne \langle \hat b_1^\dag \hat b_1^{}} \rangle $ for two unequal coherent states ($\left| \alpha  \right| \ne \left| \beta  \right|$).   
For every phase configuration, the number of photons experiencing the estimated phase is different. Therefore, the QFIs calculated from different phase configurations are different.

In particular, when $\left| \alpha  \right| = \left| \beta  \right|$, all phase configurations give the same QFI, which is found to be
\begin{align}
{\cal F} = 2{\left| \alpha  \right|^2}( {{e^{4g}} + 1} ) + {\sinh ^2}\left( {2g} \right).
\label{e13}
\end{align}
The corresponding quantum Cram\'er-Rao bound is given by
\begin{align}
\Delta \theta_{\rm{QCRB}} = \frac{1}{{\sqrt {2{{\left| \alpha  \right|}^2}\left( {{e^{4g}} + 1} \right) + {{\sinh }^2}\left( {2g} \right)} }}.
\label{e14}
\end{align}

When $\left| \alpha  \right|^2  \gg 1 $ and $e^{4g} \gg 1$, we have

\begin{align}
\Delta \theta_{\rm{QCRB}} \simeq \frac{1}{{\sqrt 2{e^{2g}}{{{\left| \alpha  \right|}} } }}.
\label{e15}
\end{align}

One can find that Eqs. (\ref{e9}) and (\ref{e15}) are the same, meaning that the phase sensitivity of our protocol approaches the quantum Cram\'er-Rao bound with the increase of $\left| \alpha  \right|^2$ and $g$.
That is, for high-intensity coherent states and high gain parameters, balanced homodyne measurement is approximately optimal strategy for our protocol.

\begin{figure}[htbp]
	\centering
	\includegraphics[width=0.48\textwidth]{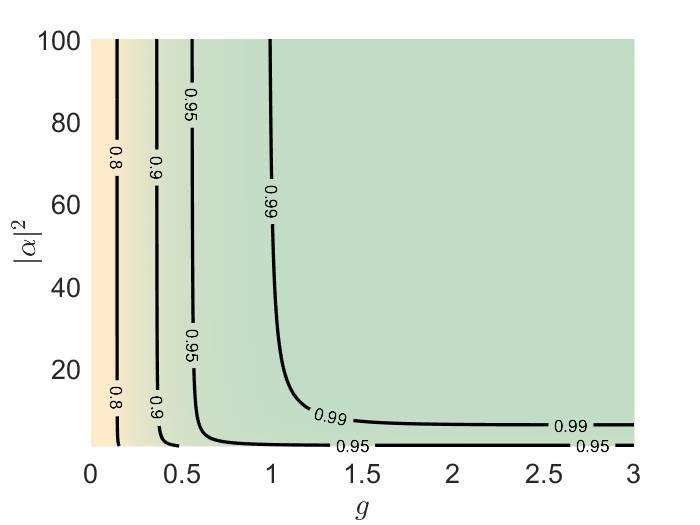}
	\caption{Ratio $\Delta \theta_{\rm{QCRB}}/\Delta \theta_{\rm{min}}$ as a function of gain parameter $g$ and mean photon number $\left| \alpha  \right|^2$ in the coherent state. The ratio is given such that a bigger ratio implies a better performance.}
	\label{sat}
\end{figure}

In Fig. \ref{sat}, we show the ratio between the ultimate phase sensitivity  $\Delta \theta_{\rm{QCRB}}$, given by quantum Cram\'er-Rao bound, and the optimal phase sensitivity $\Delta \theta_{\rm{min}}$ as a function of gain parameter $g$ and mean photon number $\left| \alpha  \right|^2$ in the coherent state. We can see that based on the current experimental techniques \cite{PhysRevLett.119.223604}, it is feasible to achieve phase sensitivity close to the one governed by the quantum Cram\'er-Rao bound.

\section{Analysis of robustness regarding photon loss and background noise}
\label{IV}

In Sec. \ref{II}, we analyzed the phase sensitivity in an ideal scenario.
In practical applications, photon loss and background noise are inevitable and they will limit the attainable phase sensitivity.
In this section, we discuss the effects of these two factors on phase sensitivity.

The process of photon loss can be modeled by adding two fictitious beam splitters with reflectivity of $L_1$ and $L_2$ in two paths \cite{PL}. 
Photons reflected by the fictitious beam splitters are regarded as photon loss, and the lossy rate in each path is equal to the reflectivity of corresponding fictitious beam splitter. 
Background noise refers to photons in the surrounding environment, which enter the interferometer through another input port of the fictitious beam splitter. 

In the optical regime, the mean photon number at room temperature is approximately $10^{-20}$ \cite{Gard2017Nearly} and the environmental mode is treated as vacuum.
However, when the measurement system is immersed in the environment with a small amount of stray light, the environmental mode is treated as a thermal state with $n_{\rm {th}} $ photons on average.
Generally, the mean photon number of background noise is much weaker than those of the input states. 
In this paper, $n_{\rm {th}} = 1 $ is used to simulated the intensity of background noise.
Related to this, $n_{\rm {th}} = 0 $ indicates that there is only photon loss, and $n_{\rm {th}} = 1$ indicates that photon loss and background noise exist simultaneously.

For a lossy scenario, the operator relationship between the input and output modes is rewritten as 	
\begin{align}
\nonumber \hat a_2^\dag  = {\kern 0.4pt} & \sqrt {\frac{{1 - {L_1}}}{2}} {e^{ - i\theta }}( {\hat a_0^\dag \cosh g + \hat b_0^{}\sinh g} ) - i\sqrt {\frac{{{L_1}}}{2}} \hat c_{}^\dag\\
&- i\sqrt {\frac{{1 - {L_2}}}{2}} ( {\hat b_0^\dag \cosh g + \hat a_0^{}\sinh g} )   - \sqrt {\frac{{{L_2}}}{2}} \hat d_{}^\dag, 
\label{e16}
\end{align}
where $\hat c_{}^\dag$ and $\hat d_{}^\dag$ represent the creation operators for environmental modes.

We can recalculate that
\begin{equation}
\langle {{{\hat X}_{\rm{L}}}} \rangle  = \sqrt {2\left( {1 - {L_1}} \right)} {e^g}\left| \alpha  \right|{\rm{cos}}\theta,
\label{e17}
\end{equation}	
and
\begin{align}
\nonumber \langle {\hat X_{\rm L}^2} \rangle  = {\kern 0.5pt} & 2{\left| \alpha  \right|^2}{e^{2g}}\left( {1 - {L_1}} \right) \cos^2 \theta + \left( {{L_1} + {L_2}} \right){n_{\rm{th}}} + 1\\
\nonumber & - \sqrt {\left( {1 - {L_1}} \right)\left( {1 - {L_2}} \right)} \sinh \left( {2g} \right)\sin \theta \\
&+ \left( {2 - {L_1} - {L_2}} \right){{\sinh }^2}g. 
\label{e18}
\end{align}	

In the limit of $L_1 = L_2 = 1$, the inputs are completely lossy and the outputs are thermal states.
Combining Eqs. (\ref{e17}) and (\ref{e18}), we have $\langle {\hat X_{\rm L}^2} \rangle - \langle {{{\hat X}_{\rm{L}}}} \rangle^2 = 2{n_{\rm{th}}} + 1 $, which matches the variance of a thermal state without information on the estimated phase.
In addition, Eq. (\ref{e17}) means that the visibility remains 100\% regardless of lossy rates.

Since the lengths of two paths are close in a practical scenario, we consider $L_1 = L_2 = L$ from now on.
The phase sensitivity is given by
\begin{align}
\Delta \theta_{\rm L}  = \frac{{\sqrt {\left( {1 - L} \right)\left[ {2{{\sinh }^2}g - \sinh \left( {2g} \right)\sin \theta } \right] + 2L{n_{\rm{th}}} + 1} }}{{\sqrt {2\left( {1 - L} \right)} {e^g}\left| {\alpha \sin \theta } \right|}}.
\label{e19}
\end{align}	

When the estimated phase sits at the optimal phase point $\theta=\pi/2$, the phase sensitivity reaches the minimum, which can be expressed as
\begin{align}
\Delta {\theta _{\rm L,\min }} = \frac{{\sqrt {1 + \gamma } }}{{\sqrt 2 {e^{2g}}\left| \alpha  \right|}}
\label{e20}	
\end{align}	
with
\begin{align}
\gamma  = \frac{L}{{1 - L}}\left( {2{n_{{\rm{th}}}} + 1} \right){e^{2g}}.
\label{e21}
\end{align}

It can be seen that $\gamma$ is the increment of measurement variance induced by photon loss and background noise.
As has been previously pointed out, phase sensitivity in an ideal scenario can outperform the shot-noise limit.
Here we analyze the maximum value of lossy rate when the phase sensitivity drops below the shot-noise limit. By solving the equation with respect to Eqs. (\ref{e10}) and (\ref{e20}), 
\begin{align}
\frac{{ {1 - L_{\rm{max}} + L_{\rm{max}}\left( {2{n_{{\rm{th}}}} + 1} \right){e^{2g}}} }}{{ {2\left( {1 - L_{\rm{max}}} \right)} {e^{4g}}\left| \alpha  \right|^2}} = \frac{1}{{ {2{{\left| \alpha  \right|}^2}{e^{2g}} + 2{{\sinh }^2}g} }},
\label{e22}
\end{align}
one can calculate the maximal tolerable lossy rate $L_{\rm{max}}$.

\begin{figure}[htbp]
	\centering
	\includegraphics[width=0.48\textwidth]{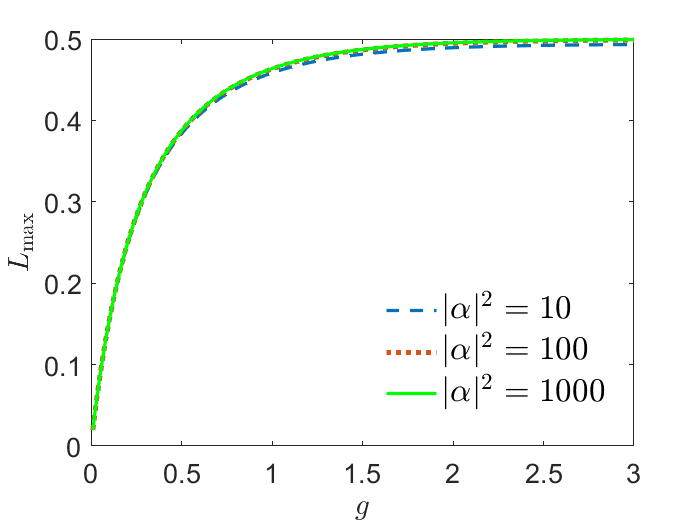}
	\caption{Maximal tolerable lossy rate as a function of gain parameter in the presence of photon loss, where $n_{\rm{th}} = 0 $.}
	\label{loss}
\end{figure}

We first consider the case of only photon loss, Fig. \ref{loss} gives the relationship between the maximal tolerable lossy rate and gain parameter with different mean photon numbers in coherent states.
The mean number of photons we use is easy to engineer under the current experimental techniques; meanwhile, the condition $|\alpha|^2 \gg 1$ is satisfied.
One can find that the effect of mean photon number on the maximal tolerable lossy rate is negligible.
With the increase of gain parameter, the maximal tolerable lossy rate increases rapidly; after about $g = 2$, this rate tends to the limit, which is equal to 50\%. 
This result indicates that increasing gain parameter can improve the robustness of our protocol.

\begin{figure}[htbp]
	\centering
	\includegraphics[width=0.48\textwidth]{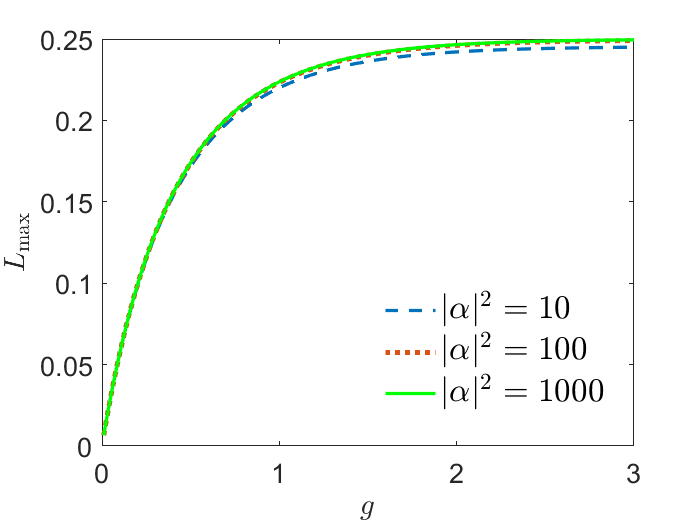}
	\caption{Maximal tolerable lossy rate as a function of gain parameter in the presence of photon loss and background noise, where $n_{\rm{th}} = 1 $.}
	\label{noise}
\end{figure}

In Fig. \ref{noise}, we show the dependence of maximal tolerable lossy rate on gain parameter in the presence of both photon loss and background noise.
Similarly, mean photon number has almost no effect on the maximal tolerable lossy rate.
The maximal tolerable lossy rate tends to the limit after about $g = 2$, and the limit is 25\%.

By comparing Fig. \ref{loss} with Fig. \ref{noise}, it can be seen that the effect of background noise is significant.
Although $n_{\rm{th}} \ll |\alpha|^2$, the maximal tolerable lossy rate decreases to half of that without background noise. 
To determine the limit of the maximal tolerable lossy rate, we can calculate the combination of Eqs. (\ref{e11}) and (\ref{e20}) as follows
\begin{align}
\frac{{1 - L_{\rm{max}} + L_{\rm{max}}\left( {2{n_{{\rm{th}}}} + 1} \right){e^{2g}}}}{{2\left( {1 - L_{\rm{max}}} \right){e^{4g}}{{\left| \alpha  \right|}^2}}} = \frac{1}{{2{{\left| \alpha  \right|}^2}{e^{2g}}}}.
\label{e23}
\end{align}

The approximate solution for Eq. (\ref{e23}) is found to be
\begin{align}
\mathop {\lim }\limits_{g \gg 1,\left| \alpha  \right| \gg 1} {L_{\max }} = \frac{1}{{2\left( {{n_{{\rm{th}}}} + 1} \right)}}.
\label{e24}
\end{align}
Obviously, we have ${L_{\max }} = 0.5$ and ${L_{\max }} = 0.25$ for $n_{\rm{th}} = 0$ and $n_{\rm{th}} = 1$, respectively.

\section{Conclusion}
\label{V}
In conclusion, we propose a new protocol for phase estimation using a modified SU(1,1) interferometer.
The second OPA in the conventional SU(1,1) interferometer is replaced by a 50-50 beam splitter.
We use two coherent states as inputs and perform balanced homodyne measurement at the output.
In an ideal scenario, the phase sensitivity is superior to the shot-noise limit by a factor of $e^g$ with coherent inputs.
Additionally, our protocol shows the potential to surpass conventional SU(1,1) interferometer in phase sensitivity. 
The improvement factor of the phase sensitivity is $\sqrt 2$ at high gain parameters and increases rapidly with the decrease of the gain parameter. 
Finally, we analyze the effects of photon loss and background noise on phase sensitivity. The results indicate that our protocol is robust against these two factors. For high gain parameters, the phase sensitivity of our protocol can outperform the shot-noise limit even when the lossy rate is relatively high. We believe that our protocol is feasible to achieve phase sensitivity close to the quantum Cram\'er-Rao bound based on current experimental techniques.

\section*{Acknowledgments} 
This work was supported by the National Natural Science Foundation of China (No. 11905160 and No. 11404040) and sponsored by Qing Lan Project of the Higher Educations of Jiangsu Province of China.

\begin{widetext}	
\appendix

\section{The exact expression for phase sensitivity calculated from classical Fisher information}
\label{A}

For many measurement strategies, error propagation formula is a simple but approximate method to calculate phase sensitivity.
An exact expression for phase sensitivity equals the inverse of the square root of the classical Fisher information, i.e., $\Delta \theta = 1/ \sqrt{{\cal F}_{\rm c}}$.
As for our protocol--Gaussian inputs and Gaussian measurement--the classical Fisher information is described as 
\begin{align}
{{\cal F}_{\rm{c}}} = \frac{{({\partial \langle {\hat X} \rangle}/{\partial \theta})^2}}{\langle {\hat X^2} \rangle - \langle {\hat X} \rangle ^2} + \frac{{2(\partial \Delta {\hat X}/{\partial \theta})^2}}{\langle {\hat X^2} \rangle - \langle {\hat X} \rangle ^2}
\end{align}
with $\Delta {\hat X} = {\sqrt{{\langle {\hat X^2} \rangle - \langle {\hat X} \rangle ^2}}}$.

One can find that the first term is related to error propagation formula. The second term at the optimal phase point, $\theta = \pi/2$, is 0 since the derivative in the numerator can be decomposed into the product of a prefactor and a cosine function, $\cos(\theta)$.
This indicates that error propagation formula and the classical Fisher information give the same result for the optimal phase sensitivity.
For other phase points, the phase sensitivity calculated from error propagation formula is greater than true value as the second term is non-negative.
These results hold true for the lossy scenario.
Overall, it is sufficient to calculate the phase sensitivity via error propagation formula, for we merely are interested in the optimal phase sensitivity of our protocol.

\section{The expectation values of measurement operators with ideal and lossy scenarios}
\label{B}
In an ideal scenario, the input-output field operator transformations of a beam splitter and those of an optical parametric amplifier are given by
\begin{align}
\hat a_2^{} &= \sqrt \eta  \hat a_1^{} e^{i\theta} + i\sqrt {1 - \eta } \hat b_1^{}, \\
\hat b_2^{} &= i\sqrt {1 - \eta } \hat a_1^{}e^{i\theta} + \sqrt \eta  \hat b_1^{}, \\
\hat a_1^{} &= \hat a_0^{}\cosh g + \hat b_0^\dag \sinh g, \\
\hat b_1^{} &= \hat b_0^{}\cosh g + \hat a_0^\dag \sinh g.
\end{align}

Regarding two equal coherent states ${\rho _A} \otimes {\rho _B}$ (${\rho _A} = {\rho _B} = \left| \alpha  \right\rangle \left\langle \alpha  \right|$), we have
\begin{align}
\langle {\hat a_2^\dag \hat a_2^\dag  + \hat a_2^{}\hat a_2^{}} \rangle  = 2{\left| \alpha  \right|^2}{e^{2g}}[ {\eta \cos \left( {2\theta } \right) - 2\sqrt {\eta \left( {1 - \eta } \right)} \sin \theta  - \left( {1 - \eta } \right)} ] - 2\sinh \left( {2g} \right)\sqrt {\eta \left( {1 - \eta } \right)} \sin \theta 
\end{align}
and
\begin{align}
\langle {\hat a_2^\dag \hat a_2^{}} \rangle  = {\left| \alpha  \right|^2}{e^{2g}}[ {1 + 2\sqrt {\eta \left( {1 - \eta } \right)} \sin \theta } ] + {\sinh ^2}g.
\end{align}

In a lossy scenario, the input-output field operator transformation of the whole interferometer is given by Eq. (\ref{e16}).
Related to this, the inputs are ${\rho _A} \otimes {\rho _B} \otimes {\rho _E} \otimes {\rho _E}$, where ${\rho _E}$ represents a thermal state with $n_{\rm{th}}$ photons on average.
Further, we have
\begin{align}
\nonumber \langle {\hat a_2^\dag \hat a_2^\dag  + \hat a_2^{}\hat a_2^{}} \rangle  = {\kern 1pt} & {\left| \alpha  \right|^2}{e^{2g}}[ { \left({1 - {L_1}}\right) \cos \left( {2\theta } \right) - 2\sqrt {\left( {1 - {L_1}} \right)\left( {1 - {L_2}} \right)} \sin \theta - \left( {1 - {L_2}} \right)} ]  \\
& - \sqrt {\left( {1 - {L_1}} \right)\left( {1 - {L_2}} \right)} \sinh \left( {2g} \right)\sin \theta
\end{align}
and
\begin{align}
\langle {\hat a_2^\dag \hat a_2^{}} \rangle  = \left( {1 - \frac{{{L_1} + {L_2}}}{2}} \right)( {{{\left| \alpha  \right|}^2}{e^{2g}} + {{\sinh }^2}g} ) + {\left| \alpha  \right|^2}{e^{2g}}\sqrt {\left( {1 - {L_1}} \right)\left( {1 - {L_2}} \right)} \sin \theta  + \frac{{{L_1} + {L_2}}}{2}{n_{\rm{th}}}.
\end{align}


\end{widetext}

%


\begin{thebibliography}{25}
\bibitem [{PL()}]{PL}%
	\BibitemOpen
	\href@noop {} {}\bibinfo {note} {In an SU(1,1) interferometer, internal loss
		has a more detrimental effect on the phase sensitivity than external loss as
		the vacuum noise is amplified by the second OPA. Owing to the fact that a
		beam splitter does not amplify vacuum noise, the effects of internal loss and
		external loss on the phase sensitivity are the same in the modified SU(1,1)
		interferometer.}\BibitemShut {Stop}%
\end{thebibliography}
\end{document}